\documentclass[preprint,prb,aps,superscriptaddress]{revtex4}
\usepackage[dvips]{graphicx}
\usepackage{color}
\usepackage{amssymb}
\usepackage{amsmath}
\usepackage{units}
\usepackage[latin1]{inputenc} 

\begin{document}

\title
{Influence of the L2$_1$ ordering degree on the magnetic
properties in Co$_2$MnSi Heusler films}

\author{O.~Gaier}
\affiliation{Fachbereich Physik and Forschungsschwerpunkt MINAS,
Technische Universit\"at Kaiserslautern,
Erwin-Schr\"odinger-Stra\ss e 56, D-67663 Kaiserslautern, Germany}

\author{J.~Hamrle}
\affiliation{Fachbereich Physik and Forschungsschwerpunkt MINAS,
Technische Universit\"at Kaiserslautern,
Erwin-Schr\"odinger-Stra\ss e 56, D-67663 Kaiserslautern, Germany}

\author{S.~J.~Hermsdoerfer}
\affiliation{Fachbereich Physik and Forschungsschwerpunkt MINAS,
Technische Universit\"at Kaiserslautern,
Erwin-Schr\"odinger-Stra\ss e 56, D-67663 Kaiserslautern, Germany}

\author{B.~Hillebrands}
\affiliation{Fachbereich Physik and Forschungsschwerpunkt MINAS,
Technische Universit\"at Kaiserslautern,
Erwin-Schr\"odinger-Stra\ss e 56, D-67663 Kaiserslautern, Germany}

\author{Y.~Sakuraba}
\affiliation{Department of Applied Physics, Graduate School of
Engineering, Tohoku University, Aoba-yama 6-6-05, Aramaki,
Aoba-ku, Sendai 980-8579, Japan}

\author{M.~Oogane}
\affiliation{Department of Applied Physics, Graduate School of
Engineering, Tohoku University, Aoba-yama 6-6-05, Aramaki,
Aoba-ku, Sendai 980-8579, Japan}

\author{Y.~Ando}
\affiliation{Department of Applied Physics, Graduate School of
Engineering, Tohoku University, Aoba-yama 6-6-05, Aramaki,
Aoba-ku, Sendai 980-8579, Japan}

\date{\today}

\begin{abstract}
We report on the influence of the improved L2$_1$ ordering degree
on the magnetic properties of Co$_2$MnSi Heusler films. Different
fractions of the L2$_1$ phase are obtained by different
post-growth annealing temperatures ranging from \unit[350]{°C} to
\unit[500]{°C}. Room temperature magneto-optical Kerr effect
measurements reveal an increase of the coercivity at an
intermediate annealing temperature of \unit[425]{°C}, which is a
fingerprint of an increased number of pinning centers at this
temperature. Furthermore, Brillouin light scattering studies show
that the improvement of the L2$_1$ order in the Co$_2$MnSi films
is correlated with a decrease of the saturation magnetization by
about \unit[9]{\%}. The exchange stiffness constant of Co$_2$MnSi,
however, increases by about \unit[8]{\%} when the L2$_1$ order is
improved. Moreover, we observe a drop of the cubic anisotropy
constant $K_1$ by a factor of 10 for an increasing amount of the
L2$_1$ phase.
\end{abstract}

\maketitle

\section{Introduction}

The performance of spin dependent devices like magnetic tunnel
junctions (MTJs) and spin valves can be considerably improved if
the conduction electrons are spin polarized. Half-metallic
ferromagnets (HMFs) are promising candidates for these
applications as they are predicted to exhibit \unit[100]{\%} spin
polarization at the Fermi level \cite{gal02, pic02}. Among
different HMFs the Heusler compound Co$_2$MnSi has attracted
particular interest due to its high Curie temperature $T_C$ of
\unit[985]{K} and a large energy gap of \unit[0.4]{eV} in the
minority spin band \cite{bro00, fuj90}.

Recently, the fabrication and the characterization of MTJs with
one electrode consisting of Co$_2$MnSi has been reported
\cite{sch04, sak05, sak06-apl88, sak06-apl89}. For example, in a
Co$_2$MnSi/Al-O/Co$_{75}$Fe$_{25}$ structure Sakuraba {\itshape et
al.} obtained a tunneling magnetoresistance (TMR) ratio of
\unit[159]{\%} at \unit[2]{K} and \unit[70]{\%} at room
temperature (RT) \cite{sak05}. According to Julliere's model, the
spin polarization at the Co$_2$MnSi interface is \unit[89]{\%} at
\unit[2]{K} \cite{sak06-apl88, sak06-apl89}. This is comparable to
a spin polarization of \unit[61]{\%} previously reported by
Schmalhorst {\itshape et al.} at \unit[10]{K}. More recently, an
increase of the TMR ratio to an extremely high value of
\unit[570]{\%} at \unit[2]{K} was achieved in a
Co$_2$MnSi/Al-O/Co$_2$MnSi structure by integrating a second
Co$_2$MnSi electrode \cite{sak06-apl88}. At RT however, no further
enhancement of the TMR ratio could be observed. This ratio was
found to be 67\% \cite{sak06-apl88} and is therefore similar to
the value obtained in the aforementioned
Co$_2$MnSi/Al-O/Co$_{75}$Fe$_{25}$ structure \cite{sak05}.

In contrast to the theoretically expected behavior the highest
value of spin polarization in Co$_2$MnSi films at RT reported so
far is only \unit[12]{\%}. This value was measured by means of
spin-resolved photoemission spectroscopy on films epitaxially
grown onto a GaAs(001) substrate \cite{wan05}. The discrepancy
between the experiment and the expected half-metallic behaviour is
mainly attributed to structural defects or site-disordering in the
Co$_2$MnSi lattice. For example, according to {\itshape ab initio}
calculations performed by Picozzi {\itshape et al.} \cite{pic04},
Co-antisite disorder leads to defect-induced states in the gap of
the minority spin band which in turn leads to a reduction of the
spin polarization.

In this work, we investigate how the magnetic properties of thin
Co$_2$MnSi films are modified when the crystal structure is
gradually improved. For this purpose, room tempera\-ture
magneto-optical Kerr effect and Brillouin light scattering studies
were performed on Co$_2$MnSi thin films with increasing fractions
of the L2$_1$ phase. Here, L2$_1$ denotes a perfectly ordered
crystal structure consisting of four interpenetrating fcc
lattices, each of which is occupied by a different kind of atoms.
In contrast, the crystal structure is referred to as B2 when Mn
and Si atoms are randomly disordered and as A2 when a random
disorder between all elements is existent. In the following, we
describe the preparation and the structural characterization of
the Co$_2$MnSi samples before presenting the results of our
magnetic investigations.

\section{Sample preparation and structural characterization}

The investigated Co$_2$MnSi films of \unit[30]{nm} thickness and
(100) orientation were epi\-taxially grown by inductively coupled
plasma-assisted magnetron sputtering onto a MgO(100) substrate
covered by a \unit[40]{nm} thick Cr(100) buffer layer. All
Co$_2$MnSi films were capped by a \unit[1.3]{nm} Al protection
layer. After the deposition, the Co$_2$MnSi films were annealed at
different annealing temperatures $T_a$ ranging from \unit[350]{°C}
to \unit[500]{°C} to provide samples with different degrees of the
L2$_1$ order.

\begin{figure}
\begin{center}
\includegraphics[width=0.6\textwidth]{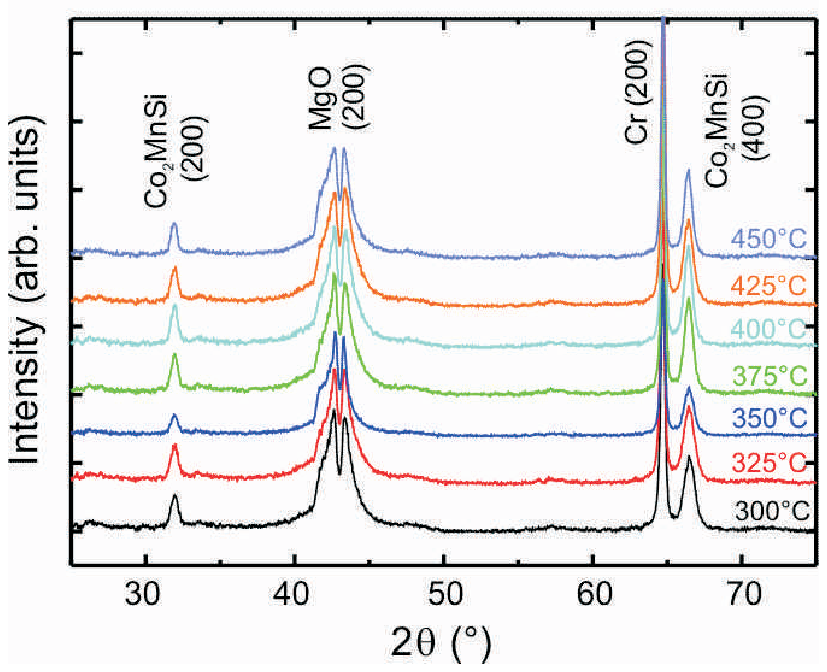}
\end{center}
\caption{%
\label{f:Bragg_scan}%
(Color online) $\theta$-2$\theta$-scans of
MgO/Cr(\unit[40]{nm})/Co$_2$MnSi(\unit[30]{nm})/Al(\unit[1.3]{nm})
films annealed at temperatures ranging from \unit[350]{°C} to
\unit[500]{°C}.}
\end{figure}

\begin{figure}
\begin{center}
\includegraphics[width=0.7\textwidth]{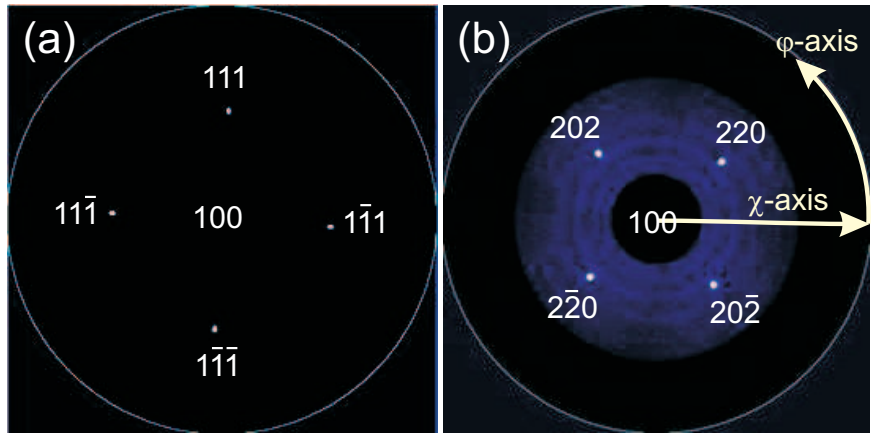}
\end{center}
\caption{%
\label{f:pole_figure}%
(Color online) X-ray pole figures for (a) (111)-planes and (b)
(220)-planes of the Co$_2$MnSi film annealed at $T_a=450$~°C.}
\end{figure}

The analysis of the crystal structure was carried out by means of
X-ray diffraction (XRD) using Cu-K$_\alpha$ radiation. The X-ray
$\theta$-2$\theta$-scans obtained from the Co$_2$MnSi samples
annealed at different temperatures $T_a$ are summarized in
Fig.~\ref{f:Bragg_scan}. For all studied films clear peaks
corresponding to the (200) and (400) reflections of Co$_2$MnSi are
observed in the specular geometry indicating perfect epitaxial
(100) growth. Furthermore, pole figure scans of all Co$_2$MnSi
samples show the presence of the (111) peak that is characteristic
for the L2$_1$ structure. An example of such a pole figure scan is
presented in Fig.~\ref{f:pole_figure}(a) for the Co$_2$MnSi film
annealed at \unit[450]{°C}. Figure~\ref{f:pole_figure}(b) shows
the pole figure of the corresponding (220) fundamental
reflections. The total integrated intensities of the (111) and
(220) reflections obtained from those pole figure scans were used
to estimate the degree of the L2$_1$ ordering in the Co$_2$MnSi
films annealed at different $T_a$. For this purpose, the
long-range order parameter $S_{L2_1}$ was calculated using the
following expression:
\begin{equation}
\label{eq:S}%
 S^2_{L2_1}=\frac{I_{obs}(111)/I_{obs}(220)}{I_{cal}(111)/I_{cal}(220)}.
\end{equation}
Here, $I_{obs}$ and $I_{cal}$ denote the peak intensities obtained
from the experiment and powder pattern simulations, respectively.
Correction factors such as the multiplicity and the film thickness
factors \cite{oka00} were taken into account when evaluating
$I_{cal}$ from the simulated intensities.

\begin{figure}
\begin{center}
\includegraphics[width=0.5\textwidth]{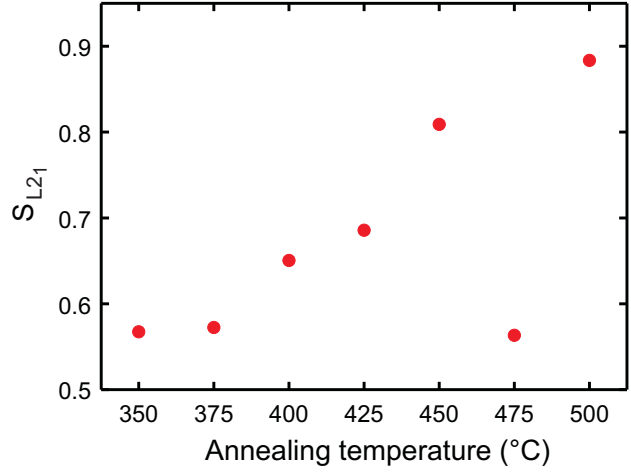}
\end{center}
\caption{%
\label{f:order_parameter}%
(Color online) Long-range order parameter $S_{L2_1}$ for
Co$_2$MnSi films annealed at different temperatures after the
deposition. The experimental error of $S_{L2_1}$ values
corresponds to the size of the data points. Note that $S_{L2_1}=1$
for a perfect order on the Mn and Si sites. }
\end{figure}

The dependence between the long-range order parameter $S_{L2_1}$
and the annealing temperature $T_a$ is presented in
Fig.~\ref{f:order_parameter}. The values of $S_{L2_1}$ increase
from approximately 0.6 to about 0.9 with increasing $T_a$ except
for the sample annealed at $T_a=$~\unit[475]{°C}. This feature is
probably related to a statistical fluctuation of the experimental
parameters during the preparation procedure. For this reason, the
Co$_2$MnSi film with $T_a=$~\unit[475]{°C} will not be considered
further in the analysis of the experimental data. $S_{L2_1}=1$ in
the case of a perfect order on the Mn and Si sites. Therefore, we
estimate that the Co$_2$MnSi film annealed at
$T_a=$~\unit[500]{°C} is \unit[80-90]{\%} L2$_1$ ordered. The
reduction of $S_{L2_1}$ with decreasing $T_a$ corresponds to the
decreasing fraction of the L2$_1$ phase inside the film.

\section{Magneto-optical Kerr effect studies}

The influence of the L2$_1$ ordering degree in Co$_2$MnSi films on
their coercivity was stu\-died by means of a standard
magneto-optical Kerr effect (MOKE) set-up in longitudinal
geometry, i.e., the external magnetic field was applied in-plane
and parallel to the plane of incidence. The sample was
investigated at room temperature by s-polarized laser light with a
wavelength of \unit[$\lambda=670$]{nm} and an angle of incidence
of 45°. MOKE hysteresis loops were measured at different sample
orientations $\alpha$ varied from 0° to 360° in steps of 1° with
$\alpha$ being the angle between the in-plane [110] direction of
the Co$_2$MnSi film and the plane of incidence.

\begin{figure}
\begin{center}
\includegraphics[width=0.9\textwidth]{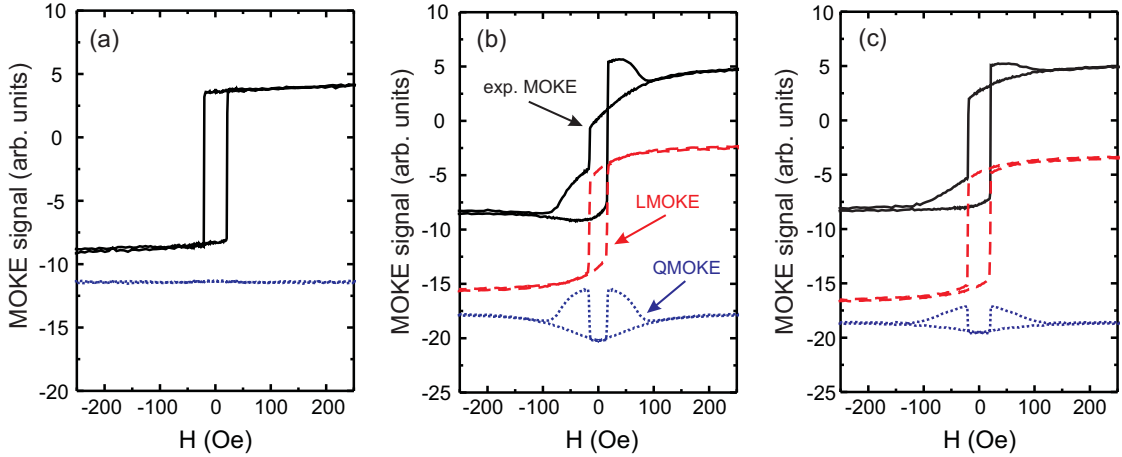}
\end{center}
\caption{%
\label{f:loops_G4}%
(Color online) Room temperature hysteresis loops measured by
magneto-optical Kerr effect (MOKE) magnetometry at sample
orientations $\alpha$ of (a) 0°, (b) 42° and (c) 45° for the
Co$_2$MnSi film annealed at \unit[425]{°C} (full line). The
symmetrization and antisymmetrization of these loops provide the
LMOKE (dashed line) and QMOKE (dotted line) contributions. }
\end{figure}

The hysteresis loops of all investigated Co$_2$MnSi films acquired
from MOKE measurements reveal a strongly asymmetric shape
depending on the sample orientation with respect to the direction
of the applied magnetic field. For example, Fig.~\ref{f:loops_G4}
shows the hysteresis curves of the Co$_2$MnSi film annealed at
$T_a=$~\unit[425]{°C} obtained for the different sample
orien\-tations $\alpha=$~0°, 42° and 45°. The observed asymmetry
results from the quadratic MOKE (QMOKE) contribution which is
proportional to quadratic terms in magnetization to the detected
signal, as already reported for various other materials
\cite{pos97, mat99, mew04, ham06quad}. By symmetrization and
antisymmetrization of experimental loops (i.e., by subtracting or
adding the loop branches for increasing and decreasing $H$ values)
it is possible to separate linear, i.e., longitudinal MOKE (LMOKE)
and QMOKE contributions \cite{mew04}, as also shown in
Fig.~\ref{f:loops_G4}.

\begin{figure}
\begin{center}
\includegraphics[width=1\textwidth]{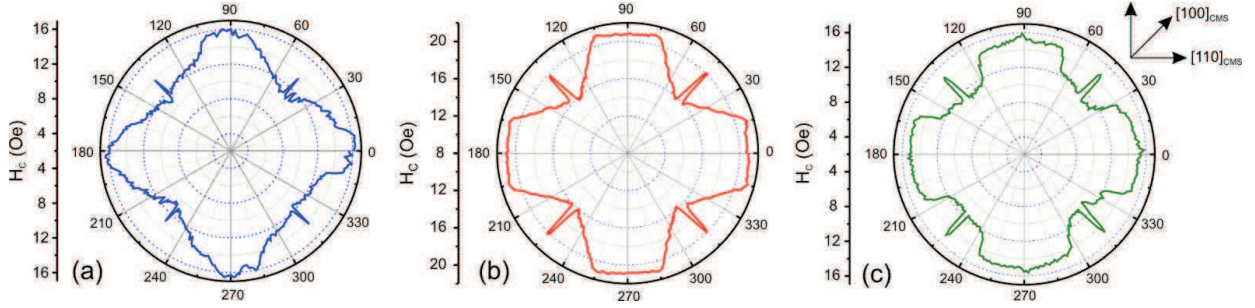}
\end{center}
\caption{%
\label{f:rotscans}%
(Color online) Angular dependence of the room temperature coercive
field $H_C$ for Co$_2$MnSi films annealed at (a) \unit[350]{°C},
(b) \unit[425]{°C} and (c) \unit[500]{°C}. $\alpha$ denotes the
angle between the external magnetic field $H$ and the [110]
crystallographic direction of the Co$_2$MnSi films. Note the
different scale for the annealing temperature of \unit[425]{°C}.}
\end{figure}

\begin{figure}
\begin{center}
\includegraphics[width=0.5\textwidth]{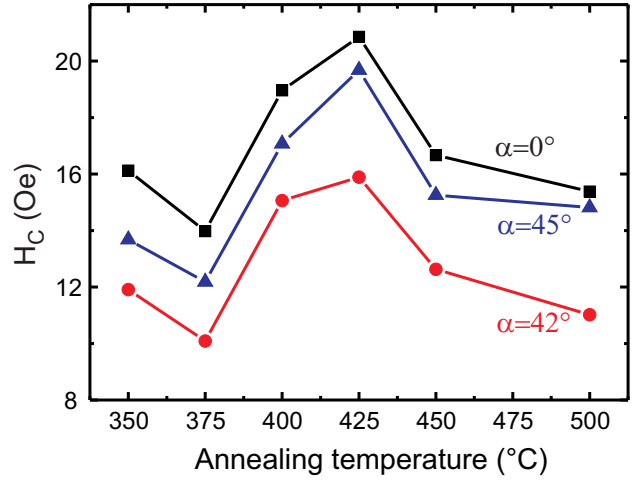}
\end{center}
\caption{%
\label{f:HcTa2}%
(Color online) Variation of the room temperature coercive field
$H_C$ with the annealing temperature for sample orientations
$\alpha=$~0°, 42° and 45°. }
\end{figure}

The LMOKE curves are used to determine the values of the coercive
field at different sample orientations and for different annealing
temperatures. The resulting polar plots (Fig.~\ref{f:rotscans})
reveal a four-fold magnetic anisotropy for all investigated
samples, reflecting the cubic symmetry of the crystal structure.
Moreover, in all cases sharp peaks in $H_C$ appear when the
applied magnetic field is aligned parallel to a hard direction.
These peaks originate from a checkerboard domain pattern which
occurs during hard axis magnetization reversal processes as
reported for \unit[80]{nm} thick Co$_2$Cr$_{0.6}$Fe$_{0.4}$Al
Heusler films \cite{ham06ccfa}. The variation of the coercivity
with $T_a$ is shown in Fig.~\ref{f:HcTa2} for three different
sample orientations. A maximum of $H_C$ is clearly visible for the
Co$_2$MnSi film annealed at \unit[425]{°C} independently from the
sample orientation $\alpha$. The observed increase of $H_C$ is
probably related to an enhanced number of pinning centers. In the
Co$_2$MnSi samples investigated here, such an increased number of
pinning centers might originate from the diffusion of Cr atoms
from the buffer layer into the Co$_2$MnSi film.

\section{Brillouin light scattering studies}

The Brillouin light scattering (BLS) technique was used to study
changes in the spin wave frequencies of Co$_2$MnSi films related
to an increasing amount of the L2$_1$ phase. All BLS spectra were
measured at room temperature at a transferred wave vector of
$q_{\|}=$~\unit[1.67]{cm$^{-1}$} using laser light with a
wavelength of $\lambda=$~\unit[532]{nm}. The magnetization vector
was oriented parallel to an in-plane easy axis direction of the
Co$_2$MnSi samples and perpendicular to the wave vector of the
incident light ($\boldsymbol{M}$$\bot$$\boldsymbol{k}$). Details
of the BLS spectrometer as well as a description of the data
acquisition have been published elsewhere \cite{hil99}.

For the geometry of the BLS set-up used here
($\boldsymbol{M}$$\bot$$\boldsymbol{k}$) two kinds of spin waves
can be excited: dipole dominated magnetostatic surface waves, also
called Damon-Eshbach modes (DE), and exchange dominated
perpendicular standing spin waves (PSSW). DE modes are
characterized by an exponential decay of the amplitude of the
dynamic magnetization over the film thickness $d$ and their
nonreciprocal behavior (i.e., propagation is possible for either
positive or negative direction of the wave vector, but not for
both). In the case of week exchange contribution and negligible
anisotropies, the wave vector dependence of DE mode frequency is
given by \cite{dem01}
\begin{equation}
\label{eq:nuDE}%
 \nu_{DE}(q_\|)=\frac{\gamma}{2\pi}\left[H(H+4\pi M_S)+(2\pi M_S)^2(1-e^{-2q_\|d})\right]^{\frac{1}{2}},
\end{equation}
where $\gamma$ is the modulus of the gyromagnetic ratio for the
electron spin, $H$ the externally applied magnetic field, $4\pi
M_S$ the saturation magnetization and $q_\|$ the in-plane wave
vector of the spin wave. PSSW modes are a superposition of two
waves propagating in opposite directions with a wave vector
$q\approx p\pi/d$ which is perpendicular to the film surface. The
positive integer $p$ denotes the quantization number of the
standing spin wave. Frequencies of the PSSW modes can
approximately be described by \cite{dem01}
{\setlength\arraycolsep{2pt}
\begin{eqnarray}
\label{eq:nuPSSW} \nu_{p} & = &
\frac{\gamma}{2\pi}\left[\left(H+\frac{2A}{M_S}q_\|^2+\frac{2A}{M_S}\left(\frac{p\pi}{d}\right)^2\right)\right.{}\nonumber\\
&&{}\left.\times\left(H+\left[\frac{2A}{M_S}+H\left(\frac{4\pi
M_S/H}{p\pi/d}\right)^2\right]q_\|^2+\frac{2A}{M_S}\left(\frac{p\pi}{d}\right)^2+4\pi
M_S\right)\right]^{\frac{1}{2}},
\end{eqnarray}}
where $A$ is the exchange stiffness constant. Note that in case of
fixed experimental conditions (i.e., $H$ and $q_\|$ are constant)
and with $\gamma$ and $d$ being unchanged, the frequency of the DE
mode (Eq.~(\ref{eq:nuDE})) only depends on the saturation
magnetization $M_S$. The frequencies of the PSSW modes
(Eq.~(\ref{eq:nuPSSW})), however, exhibit an inverse dependence on
$M_S$ and, in addition, are a function of $A$.

\begin{figure}
\begin{center}
\includegraphics[width=0.5\textwidth]{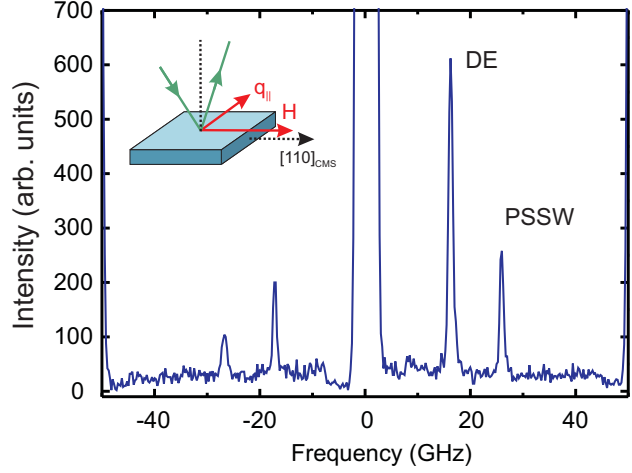}
\end{center}
\caption{%
\label{f:BLS_spectrum}%
(Color online) BLS spectrum of the Co$_2$MnSi film annealed at
\unit[400]{°C}, measured at an applied magnetic field of
$H=$~\unit[1500]{Oe} and a transferred wave vector of
$q_\|=$~\unit[1.67]{cm$^{-1}$}. $H$ was aligned parallel to the
[110] direction of the Co$_2$MnSi sample, as sketched in the
inset. }
\end{figure}

\begin{figure}
\begin{center}
\includegraphics[width=0.5\textwidth]{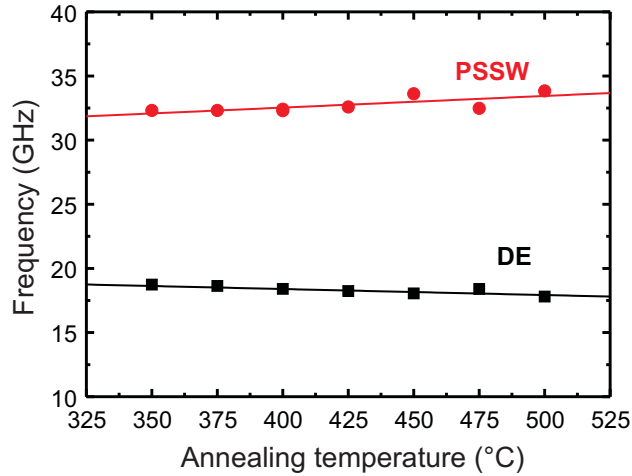}
\end{center}
\caption{%
\label{f:Peak_pos}%
(Color online) Dependence of room temperature spin wave
frequencies on the annealing temperature of the Co$_2$MnSi films.
The frequencies were determined from BLS spectra measured at an
applied magnetic field of $H=$~\unit[2000]{Oe} and a transferred
wave vector of $q_\|=$~\unit[1.67]{cm$^{-1}$}. }
\end{figure}

Figure~\ref{f:BLS_spectrum} shows a representative BLS spectrum
recorded at an external magnetic field of \unit[1500]{Oe} from the
Co$_2$MnSi film annealed at \unit[400]{°C}. In both the Stokes and
anti-Stokes regimes two peaks originating from the magnonic
scattering processes are visible. In particular, the peak at
around \unit[15]{GHz} can be attributed to the DE mode, while the
peak appearing at approximately \unit [25]{GHz} results from the
excitation of the first perpendicular standing spin wave. The
characteristic shape of the BLS spectrum remains unchanged for all
the investigated samples. A closer inspection of the spectral peak
positions, however, reveals that the frequency of the DE mode
decreases by about \unit[1]{GHz} when $T_a$ is changed from
\unit[350]{°C} to \unit[500]{°C} while the frequency of the PSSW
mode increases by about \unit[1.5]{GHz} (Fig.~\ref{f:Peak_pos}).
Equations~(\ref{eq:nuDE}) and (\ref{eq:nuPSSW}) show that the
frequency shift of the DE mode and the PSSW mode in
Fig.~(\ref{eq:nuPSSW}) is related to a decrease of the saturation
magnetization $M_S$. However, Eq.~(\ref{eq:nuPSSW}) indicates that
the increase of the PSSW mode frequency might additionally be
caused by an increase of the exchange stiffness constant $A$. From
numerical simulations using a theoretical model described in
Ref.~\cite{hil90}, we estimated that the Co$_2$MnSi film with the
smallest degree of L2$_1$ order exhibits a value of $M_S$ which is
about \unit[9]{\%} higher and an exchange stiffness constant which
is about by \unit[8]{\%} lower than the corresponding values for
the Co$_2$MnSi sample with the highest amount of the L2$_1$ phase.
Unfortunately, as BLS spectra of the investigated Co$_2$MnSi films
only showed a single PSSW mode, the numerical values of $A$ could
not be determined.

The saturation magnetization at RT for the Co$_2$MnSi film
annealed at $T_a=$~\unit[350]{°C} could be obtained by means of
vibrating sample magnetometry and was found to be
\unit[1013]{emu/cm$^3$}. This value perfectly agrees with the
theoretically predicted value of $M_S=$~\unit[5.07]{$\mu_B$/f.u.}
(i.e., \unit[1040]{emu/cm$^3$}) for the bulk Co$_2$MnSi with
L2$_1$ order \cite{web71}. Taking into account the decrease of
$M_S$ by \unit[9]{\%} obtained from numerical simulations, the
experimentally obtained value of $M_S$ is still in a good
agreement with its theoretical counterpart.

The study of spin wave frequencies in dependence of the in-plane
sample orientation allows for the determination of anisotropy
constants of thin magnetic films. We performed such BLS
spectroscopy measurements for the Co$_2$MnSi films annealed at
350, 375, 400, 450 and \unit[500]{°C}. All spectra were recorded
at a transferred wave vector of $q_{\|}=$~\unit[1.67]{cm$^{-1}$}
and an external magnetic field of $H=$~\unit[300]{Oe}, which was
large enough to saturate the sample. The angle $\alpha$ between
$H$ and the [110] easy axis direction of the Co$_2$MnSi films was
varied from \unit[0]{°} to \unit[180]{°} in \unit[15]{°} steps by
rotating the sample. Due to the cubic symmetry of the crystal
structure, the variation of $\alpha$ in the whole range from
\unit[0]{°} to \unit[360]{°} was not required.

\begin{figure}
\begin{center}
\includegraphics[width=0.5\textwidth]{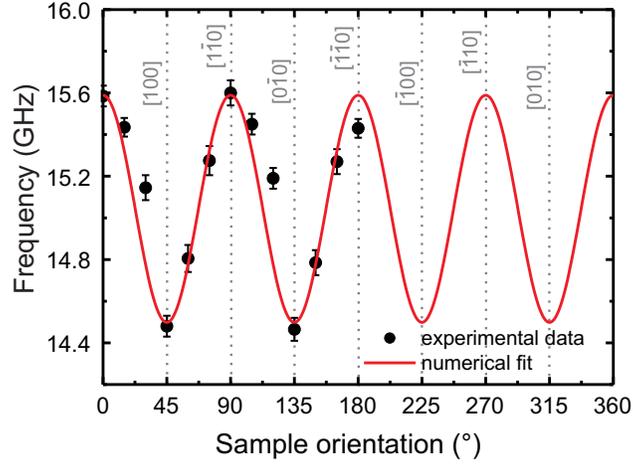}
\end{center}
\caption{%
\label{f:DE_anisotropy}%
(Color online) Frequency of the DE mode as function of the angle
between the external magnetic field $H=$~\unit[300]{Oe} and the
[110] easy axis direction for the Co$_2$MnSi film annealed at
\unit[375]{°C}. }
\end{figure}

\begin{figure}
\begin{center}
\includegraphics[width=0.5\textwidth]{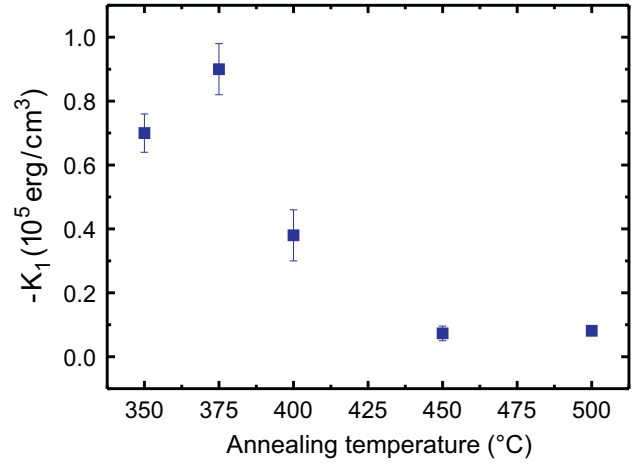}
\end{center}
\caption{%
\label{f:K1}%
(Color online) Room temperature volume anisotropy constant $K_1$
for Co$_2$MnSi films annealed at different temperatures $T_a$. }
\end{figure}

Figure~\ref{f:DE_anisotropy} shows the frequency of the DE mode as
function of the external magnetic field $H$ for the Co$_2$MnSi
film annealed at $T_a=$~\unit[375]{°C}. The sample exhibits a
clear four-fold magnetic anisotropy which perfectly agrees with
the results obtained from the MOKE investigations presented in the
previous section. Maxima of the surface spin wave frequencies at
$\alpha=n\times90°$, with $n$ being an integer, show that the
[110] directions are easy axis directions. Furthermore, from the
minima of the DE frequencies at $(n+1/2)\times90°$ we deduce that
the [100] directions are hard axis directions. Fitting the
theoretical curve obtained by means of the model described in
Ref.~\cite{hil90} to the experimental data, we determined the
cubic volume anisotropy constant $K_1$. The values of $K_1$ for
samples annealed at different temperatures are shown in
Fig.~\ref{f:K1}. The maximal value of
$K_1=-9\cdot10^5$~{erg/cm$^3$} was found for
$T_a=$~\unit[375]{°C}. The increase of $T_a$ leads to a drop of
the volume anisotropy constant by a factor of 10, which might be
related to the improvement of the L2$_1$ order. However, other
factors such as diffusion of Cr into the Co$_2$MnSi layer might
play a role as well.

\section{Summary}

The influence of the L2$_1$ ordering degree on the magnetic
properties of the Co$_2$MnSi Heusler compound was investigated.
Different amounts of the L2$_1$ phase were realized by different
post-growth annealing temperatures ranging from \unit[350]{°C} to
\unit[500]{°C}. Summa\-rizing the results of the magnetic
characterization using MOKE and BLS we reveal that an increased
L2$_1$ ordering degree of the Co$_2$MnSi films is correlated with
a reduction of the saturation magnetization of about \unit[9]{\%},
an increase of the exchange stiffness constant of \unit[8]{\%} and
a drop of the cubic anisotropy constant $K_1$ by a factor of 10.

\section{Acknowledgment}

The project was financially supported by the Research Unit 559
\emph{"New materials with high spin polarization"} funded by the
Deutsche Forschungsgemeinschaft and by the NEDO International
Joint Research Grant Programm 2004/T093. We would like to thank
M.~Jourdan, H.~Schneider and  H.~Schulthei{\ss} for fruitful
discussions and technical advice.

\newpage

\newpage

\end{document}